\documentclass[preprint,12pt]{elsarticle}

\usepackage{hyperref}
\usepackage{cleveref}

\usepackage[utf8]{inputenc}
\usepackage{cases}
\usepackage{amsthm}
\usepackage{amsfonts}
\usepackage{multicol}
\usepackage{graphicx}
\usepackage{epsfig}
\usepackage{pdfpages}
\usepackage{amssymb}

\usepackage{grffile}        % Omite o nome do arquivo de uma figura
\usepackage{caption}
\usepackage{subcaption}

\DeclareGraphicsExtensions{.jpg}

\journal{Communications in Nonlinear Science and Numerical Simulation}

\begin{document}

\newtheorem{conjectura}{Conjecture}
\newtheorem{definition}{Definition}
\newtheorem{lema}{Lema}
\newtheorem{theorem}{Theorem}
\newtheorem{proposition}{Proposition}

\begin{frontmatter}

\title{Classical solution for the linear sigma model}

\author{Danilo V. Ruy}

\address{Departamento de Física, Universidade Federal da Paraíba, 58051-970 João Pessoa PB, Brazil}
\ead{daniloruy@ift.unesp.br}

\begin{abstract}

In this paper, the linear sigma model is studied using a method for finding analytical solutions based on Padé approximants. Using the solutions of two and three traveling waves in 1+3 dimensions we found, we are able to show a solution that is valid for an arbitrary number of bosons and traveling waves.

\end{abstract}


\end{frontmatter}

%%%%%%%%%%%%%%%%%%%%%%%%%%%%%%%%%%%%%%%%%%%%%%%%%%%%%%%%%%%%%%%%%%%

\section{Introduction}

Field theory is a well established way to describe interactions among particles, but usually it is very challenging to find analytical classical solutions in 1+3 dimensions. In this paper, we will study the bosonic interaction of the linear sigma model, which was proposed in \cite{Gell-Mann} as an model to describe the pion-nucleon interaction. The model we are considering possess $N_\phi$ bosonic fields whose Lagrangian is given by
\[
{\cal L}={1\over2}(\partial_\mu{\bf \Phi})^2-{1\over2}m^2({\bf \Phi})^2-{\lambda\over4}(({\bf \Phi})^2)^2
\]
where $({\bf\Phi})^2\equiv {\bf\Phi}\cdot{\bf\Phi}$ and ${\bf \Phi}=(\phi_1,\phi_2,...,\phi_{N_\phi})$. The Euler-Lagrange equation for this Lagrangian yields the system of equations
\begin{equation}\label{linear sima model eq}
\phi_{i,tt}-\nabla\phi_i+m^2\phi_i+\lambda\biggl(\phi_i^3+\phi_i\sum_{j=1}^{N_{\phi}-1}\phi_{i+j}^2\biggr)=0,    \hspace{1 cm}   i=1,...,N_{\phi},
\end{equation}
where  we use the notation $\phi_i=\phi_{i+N_{\phi}}$ and the metric $(+---)$. The purpose of this paper is to seek classical solutions for this system in the form of traveling waves. In order to achieve that, we will employ  the algorithm based on Padé approximants presented in \cite{Ruy,Ruy lambda phi 4}. This method has the advantage of needing less undetermined constants than other methods in the literature \cite{Parkes,Malfliet,Kudryashov01,kudryashov,Parkes2,Vitanov1,Vitanov2,Vitanov3,Biswas1,Biswas2,He,Wu,Ma1,Ma2,Wang,Ebadi,Lee}, such that we can earn efficiency in the processing of data.

We briefly recall the algorithm from \cite{Ruy lambda phi 4} in section (\ref{method}) and we apply it to the linear sigma model in section (\ref{application}). Using the solutions for two and three traveling waves we found, we are able to show a solution for an arbitrary number of traveling waves and bosons in section (\ref{general case}). Section (\ref{illustration}) is devoted to show a graphical illustration of the analytical solution we found. Finally, in section \ref{comparison}, we compare the Padé approximant approach with the multiple exp-method in order to see the benefits of each method.

\section{Review of the algorithm}\label{method}

Consider a system of $N_e$ equation in $D$ dimensions
\begin{equation}\label{system E}
E_k(x^\mu,\phi_i,\partial_\mu\phi_i,...;{\cal S}_0)=0,    \hspace{1 cm}   k=1,...,N_e.
\end{equation}
where ${\cal S}_0$ is the set of all parameters of the model. Now suppose there is at least one solution of this system which can be written  in terms of $N_\rho$ functions we know the first derivative, i.e.
\begin{eqnarray*}
 \phi_i(x^\mu)=\hat{\phi}_i(\rho_1,...,\rho_{N_\rho})  ,  &  \hspace{1 cm} &     i=1,...,N_e    \\
  \partial_\mu\rho_k={\cal F}_{\mu,k}(\rho_1,...,\rho_{N_\rho};{\cal S}_1),    &  \hspace{1 cm} &  \mu=1,...,D    \hspace{1 cm}     k=1,...,N_\rho
\end{eqnarray*}
where ${\cal F}_{\mu,k}(\rho_1,...,\rho_{N_\rho};{\cal S}_1)$ is determined by the functions we choose for $\rho_k$ and ${\cal S}_1$ is the set of constants introduced by these functions. This ansatz transforms the system (\ref{system E}) as
\begin{equation}\label{eq E}
E_k(x^\mu,\phi_i,\partial_\mu\phi_i,...;{\cal S}_0)=\hat{E}_k(\rho_k,\hat{\phi}_i,\partial_k\hat{\phi}_i,...;{\cal S}_0\times{\cal S}_1)=0,    \hspace{1 cm}   k=1,...,N_e.
\end{equation}

Suppose there is a solution that is regular when $\rho_k\to 0$ such that we can calculate the multivariate Taylor expansion at $\rho=0$, i. e.
\begin{equation}\label{Taylor teorico}
\hat{\phi}_i(\rho)=\sum_{j_1=0}^\infty\sum_{j_2=0}^\infty ...\sum_{j_{N_\rho}=0}^\infty c^{(i)}_{j_1,j_2,...,j_{N_\rho}}\prod_{d=1}^{N_\rho} \rho_{d}^{j_d}  ,    \hspace{1 cm}   i=1,...,N_e
\end{equation}
In the expansion (\ref{Taylor teorico}), we may also impose a set of constraints $\psi_i({\cal S}_0\times{\cal S}_1)=0$ on the undetermined constants such that the series does not truncate in the vacuum case. We will call ${\cal S}_2$ the set of all undetermined $c^{(i)}_{j_1,j_2,...,j_{N_\rho}}$ and ${\cal S}={\cal S}_0\times{\cal S}_1\times{\cal S}_2$ the set of all constants that remains undetermined. Let us do the scaling transformation $\rho\to\xi\rho$ such that we can rearrange the Taylor expansion as
\[
\hat{\phi}_i(\xi\rho)= \sum_{n=0}^{L_i+M_i}a_{i;n}(\rho) \xi^n+{\cal O}(\xi^{L_i+M_i+1})  ,    \hspace{1 cm}   i=1,...,N_e  ,
\]
\[
a_{i;n}(\rho)=\sum_{j_1=0}^{n}\sum_{j_2=0}^{n-j_1} ...\sum_{j_{{N_\rho}-1}=0}^{n-\sum_{r=1}^{{N_\rho}-2} j_{r}} c^{(i)}_{j_1,j_2,...,j_{{N_\rho}-1},n-\sum_{r=1}^{{N_\rho}-1} j_{r}}\biggl(\prod_{d=1}^{{N_\rho}-1} \rho_{d}^{j_d}\biggr) \rho_{N_\rho}^{n-\sum_{r=1}^{{N_\rho}-1} j_{r}}.
\]
The coefficients of the expansion (\ref{Taylor teorico}) need be calculated until we determine all $a_{i;n}(\rho)$ up to order $L_i+M_i$. Now, for a chosen $L_i$ and $M_i$, we can calculate the Padé approximant for the field using $\xi$ as variable, i.e.
\begin{equation}\label{aproximado}
\hat{\phi}_i(\xi \rho )={P_{\rho,L_i}^{(i)}(\xi;{\cal S})\over Q_{\rho,M_i}^{(i)}(\xi;{\cal S})}+{\cal O}(\xi^{L_i+M_i+1}).
\end{equation}
The Padé approximants consist in an approximation of a function as a ratio of two polynomials given by
\[
[L_i/M_i]_\rho^{(i)}(\xi;{\cal S})\equiv{P_{\rho,L_i}^{(i)}(\xi;{\cal S})\over Q_{\rho,M_i}^{(i)}(\xi;{\cal S})}={\sum_{j=0}^{L_i} p_j^{(i)}(\rho;{\cal S})\xi^j\over 1+\sum_{j=1}^{M_i} q_j^{(i)}(\rho;{\cal S})\xi^j}
\]
whose coefficient are calculated by the system of equations
\[
\sum_{r+s=j}a_{i;r}(\rho)q_s^{(i)}(\rho;{\cal S})-p_j^{(i)}(\rho;{\cal S})=0,   \hspace{2 cm}   j=0,1,...,L_i+M_i.
\]
Now we would like to find the subset $\hat{{\cal S}}\subset{\cal S}$ which makes the approximation (\ref{aproximado}) become an exact solution for the system (\ref{system E}) when $\xi=1$, i.e.
\begin{equation}\label{sol exata}
\hat{\phi}_i(\rho)={P_{\rho,L_i}^{(i)}(\xi;\hat{{\cal S}})\over Q_{\rho,M_i}^{(i)}(\xi;\hat{{\cal S}})}\biggl|_{\xi=1}.
\end{equation}
Substituting (\ref{sol exata}) in (\ref{eq E}), we can rewrite the system as
\[
\hat{E}_k(\rho_k,\hat{\phi}_i,\partial_k\hat{\phi}_i,...;{\cal S}_0\times{\cal S}_1)={\sum_{n=0}^{\Lambda}\hat{E}_{k;n}(\hat{S})  \over D_k(\hat{S}) }=0,
\]
\begin{equation}
\hat{E}_{k;n}(\hat{S})=\sum_{j_1=0}^{n}\sum_{j_2=0}^{n-j_1} ...\sum_{j_{{N_\rho}-1}=0}^{n-\sum_{r=1}^{{N_\rho}-2} j_{r}} \tilde{E}_{k;j_1,j_2,...,j_{{N_\rho}-1},n-\sum_{r=1}^{{N_\rho}-1} j_{r}}\biggl(\prod_{d=1}^{{N_\rho}-1} \rho_{d}^{j_d}\biggr) \rho_{N_\rho}^{n-\sum_{r=1}^{{N_\rho}-1} j_{r}}.
\end{equation}
The coefficients for all powers of $\rho_k$ yield the set of algebraic equations
\begin{subequations}
\begin{eqnarray}\label{cond teorico}
\tilde{E}_{k;j_1,j_2,...,j_{{N_\rho}}}(\hat{S}) = 0  ,  &&       k=1,...,N_e  ,   \hspace{0.4 cm}  n=0,...,\Lambda  ,   \hspace{0.4 cm} j_l=0,...,n-\sum_{r=1}^{l-1}j_r  \\
 D_k(\hat{S}) \neq 0  ,   &&    k=1,...,N_e 
\end{eqnarray}
\end{subequations}
whose solution will determine the subset of constants $\hat{{\cal S}}$. After determine $\hat{{\cal S}}$, we substitute its elements back into (\ref{sol exata}).  Solutions of (\ref{cond teorico}) which simplify (\ref{sol exata}) to a vacuum solution will be omitted through the paper.

\section{Application to the linear sigma model}\label{application}

In this paper, let us consider a ansatz with $N_{\rho}$ traveling waves  that obey a structure of exponential functions, i.e.
\begin{equation}\label{general ansatz}
\phi_i(x^\mu)=\hat{\phi}_i(\rho_1,...,\rho_{N_{\rho}}),    \hspace{1.0 cm}   \partial_\mu\rho_{j}=k_{j\mu}\rho_j,    \hspace{1 cm}   i=1,...,N_{\phi},    \hspace{1 cm}   j=1,...,N_{\rho}.
\end{equation}

Substituting the ansatz (\ref{general ansatz}) into system (\ref{linear sima model eq}), we map the original system in 1+3 dimensions to a system in $N_\rho$ dimensions with the form
\begin{equation}\label{system rho}
\sum_{p,q=1}^{N_{\rho}}k_{p\mu}k_q^\mu\biggl(\rho_p\rho_q\hat{\phi}_{i,\rho_p\rho_q}+\delta_{pq}\rho_p\hat{\phi}_{i,\rho_p}\biggr)+m^2\hat{\phi}_i+\lambda\biggl(\hat{\phi}_i^3+\hat{\phi}_i\sum_{j=1}^{N_{\phi}-1}\hat{\phi}_{i+j}^2\biggr)=0,    \hspace{1 cm}   i=1,...,N_{\phi}
\end{equation}

In the subsections below, we will consider the cases with $N_\phi=N_\rho=2$ and $N_\phi=N_\rho=3$. With these solution, we are able to show a solution for a general case with an arbitrary number of fields and traveling waves in subsection \ref{general case}. In both cases, we will employ Padé approximants with $L_i=M_i=2$ for all $i$.

\subsection{Case with $N_{\phi}=2$ and $N_{\rho}=2$} \label{subsection N2}

In order to calculate a nonzero Taylor expansion, we will impose the constraints
\[\
k_{1\mu}k_1^{\mu}+m^2=0,   \hspace{1 cm}      k_{2\mu}k_2^{\mu}+m^2=0.
\]
With these constraints, we can eliminate $k_{10}$ and $k_{20}$. However, for simplify the calculation, we will eliminate only the quadratic terms of $k_{10}$ and $k_{20}$ at this point.
Let us consider the Taylor expansions for the fields $\hat{\phi}_1=\sum_{i,j=0}c_{ij}\rho_1^i\rho_2^j$ and $\hat{\phi}_2=\sum_{i,j=0}d_{ij}\rho_1^i\rho_2^j$, such that after the rescaling $\rho\to\xi\rho$ we have
\small
\begin{eqnarray*}
\hat{\phi}_1(\xi \rho ) &=&  \xi(c_{10}\rho_1+c_{01}\rho_2)+\xi^3\biggl({c_{10}(c_{10}^2+d_{10}^2)\lambda \rho_1^3\over 8m^2}-{(3c_{01}c_{10}^2+2c_{10}d_{10}d_{01}+c_{01}d_{10}^2)\lambda \rho_1^2\rho_2\over 4(k_{10}k_{20}-k_{11}k_{21}-k_{12}k_{22}-k_{13}k_{23}-m^2)}   \\
 &-&  {(3c_{01}^2c_{10}+2c_{01}d_{10}d_{01}+c_{10}d_{01}^2)\lambda \rho_1\rho_2^2\over 4(k_{10}k_{20}-k_{11}k_{21}-k_{12}k_{22}-k_{13}k_{23}-m^2)}+{c_{01}(c_{01}^2+d_{01}^2)\lambda \rho_2^3\over 8m^2}      \biggr)+O(\xi^5) \\
&& \\
\hat{\phi}_2(\xi \rho ) &=&  \xi(d_{10}\rho_1+d_{01}\rho_2)+\xi^3\biggl({d_{10}(c_{10}^2+d_{10}^2)\lambda \rho_1^3\over 8m^2}-{(3d_{01}d_{10}^2+2d_{10}c_{10}c_{01}+d_{01}c_{10}^2)\lambda \rho_1^2\rho_2\over 4(k_{10}k_{20}-k_{11}k_{21}-k_{12}k_{22}-k_{13}k_{23}-m^2)}   \\
 &-&  {(3d_{01}^2d_{10}+2d_{01}c_{10}c_{01}+d_{10}c_{01}^2)\lambda \rho_1\rho_2^2\over 4(k_{10}k_{20}-k_{11}k_{21}-k_{12}k_{22}-k_{13}k_{23}-m^2)}+{d_{01}(c_{01}^2+d_{01}^2)\lambda \rho_2^3\over 8m^2}      \biggr)+O(\xi^5)
\end{eqnarray*}
\normalsize
Considering the expansion until  order 4 for the auxiliary parameter $\xi$, we can calculate the Padé approximants $[2/2]_\rho^{(i)}(\xi;{\cal S})$ as 
\begin{eqnarray*}
{P_{\rho,2}^{(1)}(\xi;{\cal S})\over Q_{\rho,2}^{(1)}(\xi;{\cal S})}\biggl|_{\xi=1} &=&   8m^2(-k_{10}k_{20}+k_{11}k_{21}+k_{12}k_{22}+k_{13}k_{23}+m^2)(c_{10}\rho_1+c_{01}\rho_2)^2\biggl/   \\
&&\biggl[ c_{10}(k_{10}k_{20}-k_{11}k_{21}-k_{12}k_{22}-k_{13}k_{23}-m^2)(-8m^2+(c_{10}^2+d_{10}^2)\lambda\rho_1^2)\rho_1  \\
&+&2m^2(4c_{01}(-k_{10}k_{20}+k_{11}k_{21}+k_{12}k_{22}+k_{13}k_{23}+m^2)-(2c_{10}d_{01}d_{10}   \\
&+& c_{01}(3c_{10}^2+d_{10}^2))\lambda\rho_1^2)\rho_2-2(3c_{01}^2c_{10}+c_{10}d_{01}^2+2c_{01}d_{01}d_{10})m^2\lambda\rho_1\rho_2^2 \\
&-&c_{01}(c_{01}^2+d_{01}^2)(-k_{10}k_{20}+k_{11}k_{21}+k_{12}k_{22}+k_{13}k_{23}+m^2)\lambda\rho_2^3\biggr]  \\
&& \\
{P_{\rho,2}^{(2)}(\xi;{\cal S})\over Q_{\rho,2}^{(2)}(\xi;{\cal S})}\biggl|_{\xi=1} &=&   8m^2(-k_{10}k_{20}+k_{11}k_{21}+k_{12}k_{22}+k_{13}k_{23}+m^2)(d_{10}\rho_1+d_{01}\rho_2)^2\biggl/   \\
&&\biggl[ d_{10}(k_{10}k_{20}-k_{11}k_{21}-k_{12}k_{22}-k_{13}k_{23}-m^2)(-8m^2+(c_{10}^2+d_{10}^2)\lambda\rho_1^2)\rho_1  \\
&+&2m^2(4d_{01}(-k_{10}k_{20}+k_{11}k_{21}+k_{12}k_{22}+k_{13}k_{23}+m^2)-(2d_{10}c_{01}c_{10}   \\
&+& d_{01}(3d_{10}^2+c_{10}^2))\lambda\rho_1^2)\rho_2-2(3d_{01}^2d_{10}+d_{10}c_{01}^2+2d_{01}c_{01}c_{10})m^2\lambda\rho_1\rho_2^2 \\
&-&d_{01}(c_{01}^2+d_{01}^2)(-k_{10}k_{20}+k_{11}k_{21}+k_{12}k_{22}+k_{13}k_{23}+m^2)\lambda\rho_2^3\biggr]
\end{eqnarray*}
where ${\cal S}=\{m,\lambda,k_{10},k_{11},k_{12},k_{13},k_{20},k_{21},k_{22},k_{23},c_{10},c_{01},d_{10},d_{01}\}$. Following step (\ref{sol exata}) of the algorithm, we substitute these Padé approximants into the system (\ref{system rho}) with $N_\phi=N_\rho=2$ and it yields a set of 66 algebraic equations that need be solved in order to determine the subset $\hat{{\cal S}}\subset{\cal S}$. We will not write the 66 algebraic equations, but its non-trivial solutions together with the constraints imposed on the Taylor expansion can be summarized by the following conditions on the constants $k_{i\mu}$:
\begin{equation}\label{condition 2 waves}
k_{1\mu}k_1^\mu+m^2=0,    \hspace{1 cm} k_{2\mu}k_2^\mu+m^2=0,    \hspace{1 cm}  k_{1\mu}k_2^\mu+m^2=0
\end{equation}

Finally, substituting this relations on the ansatz
\[
\hat{\phi}_1={P_{\rho,2}^{(1)}(\xi;\hat{{\cal S}})\over Q_{\rho,2}^{(1)}(\xi;\hat{{\cal S}})}\biggl|_{\xi=1} ,    \hspace{1 cm}    \hat{\phi}_2={P_{\rho,2}^{(2)}(\xi;\hat{{\cal S}})\over Q_{\rho,2}^{(2)}(\xi;\hat{{\cal S}})}\biggl|_{\xi=1} , 
\]
we have the solution
\[
\hat{\phi}_1={8m^2(c_{10}\rho_1+c_{01}\rho_2)\over 8m^2-\lambda[(c_{10}^2+d_{10}^2)\rho_1^2+2(c_{10}c_{01}+d_{10}d_{01})\rho_1\rho_2+(c_{01}^2+d_{01}^2)\rho_2^2]}
\]
\[
\hat{\phi}_2={8m^2(d_{10}\rho_1+d_{01}\rho_2)\over 8m^2-\lambda[(c_{10}^2+d_{10}^2)\rho_1^2+2(c_{10}c_{01}+d_{10}d_{01})\rho_1\rho_2+(c_{01}^2+d_{01}^2)\rho_2^2]}
\]

\subsection{Case with $N_{\phi}=3$ and $N_{\rho}=3$}\label{subsection N3}

In this section, we will exploit the linear sigma model with $N_\phi=N_\rho=3$. As we are dealing with a system of 3 field in 1+3 dimensions, the processing of the algebraic data increases considerably. So we will extend the information we found for $N_\phi=N_\rho=2$ in order to simplify the calculation and consider the constraints 
\begin{eqnarray}
&& k_{1\mu}k_1^\mu+m^2=0,    \hspace{1 cm}   k_{2\mu}k_2^\mu+m^2=0,    \hspace{1 cm} k_{3\mu}k_3^\mu+m^2=0,  \label{constraint 3a}     \\
&& k_{1\mu}k_2^\mu+m^2=0,    \hspace{1 cm}    k_{1\mu}k_3^\mu+m^2=0,    \hspace{1 cm}    k_{2\mu}k_3^\mu+m^2=0   \label{constraint 3b}
\end{eqnarray}
on the Taylor expansions $\hat{\phi}_1=\sum_{i,j,k=0}c_{ijk}\rho_1^i\rho_2^j\rho_3^k$, $\hat{\phi}_2=\sum_{i,j,k=0}d_{ijk}\rho_1^i\rho_2^j\rho_3^k$ and $\hat{\phi}_3=\sum_{i,j,k=0}e_{ijk}\rho_1^i\rho_2^j\rho_3^k$. Although we only need constraint (\ref{constraint 3a}) to avoid the vacuum solution, the constraint (\ref{constraint 3b}) simplify considerably the calculation as we will see. The expansions are
\small
\begin{eqnarray*}
\hat{\phi}_1(\xi \rho ) &=&  \xi(c_{100}\rho_1+c_{010}\rho_2+c_{001}\rho_3)+{\xi^3\lambda\over 8m^2}\biggl[(c_{100}\rho_1+c_{010}\rho_2+c_{001}\rho_3)((c_{100}^2+d_{100}^2+e_{100}^2)\rho_1^2 \\
&+& (c_{010}^2+d_{010}^2+e_{010}^2)\rho_2^2+(c_{001}^2+d_{001}^2+e_{001}^2)\rho_3^2+2(c_{100}c_{010}+d_{100}d_{010}+e_{100}e_{010})\rho_1\rho_2 \\
&+& 2(c_{100}c_{001}+d_{100}d_{001}+e_{100}e_{001})\rho_1\rho_3 + 2(c_{010}c_{001}+d_{010}d_{001}+e_{010}e_{001})\rho_2\rho_3\biggr]+O(\xi^5)  \\
\hat{\phi}_2(\xi \rho ) &=&  \xi(d_{100}\rho_1+d_{010}\rho_2+d_{001}\rho_3)+{\xi^3\lambda\over 8m^2}\biggl[(d_{100}\rho_1+d_{010}\rho_2+d_{001}\rho_3)((c_{100}^2+d_{100}^2+e_{100}^2)\rho_1^2 \\
&+& (c_{010}^2+d_{010}^2+e_{010}^2)\rho_2^2+(c_{001}^2+d_{001}^2+e_{001}^2)\rho_3^2+2(c_{100}c_{010}+d_{100}d_{010}+e_{100}e_{010})\rho_1\rho_2 \\
&+& 2(c_{100}c_{001}+d_{100}d_{001}+e_{100}e_{001})\rho_1\rho_3 + 2(c_{010}c_{001}+d_{010}d_{001}+e_{010}e_{001})\rho_2\rho_3\biggr]+O(\xi^5)  \\
\hat{\phi}_3(\xi \rho ) &=&  \xi(e_{100}\rho_1+e_{010}\rho_2+e_{001}\rho_3)+{\xi^3\lambda\over 8m^2}\biggl[(e_{100}\rho_1+e_{010}\rho_2+e_{001}\rho_3)((c_{100}^2+d_{100}^2+e_{100}^2)\rho_1^2 \\
&+& (c_{010}^2+d_{010}^2+e_{010}^2)\rho_2^2+(c_{001}^2+d_{001}^2+e_{001}^2)\rho_3^2+2(c_{100}c_{010}+d_{100}d_{010}+e_{100}e_{010})\rho_1\rho_2 \\
&+& 2(c_{100}c_{001}+d_{100}d_{001}+e_{100}e_{001})\rho_1\rho_3 + 2(c_{010}c_{001}+d_{010}d_{001}+e_{010}e_{001})\rho_2\rho_3\biggr] +O(\xi^5)
\end{eqnarray*}
\normalsize

Considering the expansion until  order 4 for the auxiliary parameter $\xi$, we can calculate the ansatz $\hat{\phi}_i=[2/2]_\rho^{(i)}(\xi;{\cal S})$ as 
\small
\begin{eqnarray*}
\hat{\phi}_1&=&8m^2(c_{100}\rho_1+c_{010}\rho_2+c_{001}\rho_3)\biggl/\biggl[ 8m^2-\lambda[(c_{100}^2+d_{100}^2+e_{100}^2)\rho_1^2 \\
&&+(c_{010}^2+d_{010}^2+e_{010}^2)\rho_2^2+(c_{001}^2+d_{001}^2+e_{001}^2)\rho_3^2+2(c_{100}c_{010}+d_{100}d_{010}+e_{100}e_{010})\rho_1\rho_2 \\
&&+2(c_{100}c_{001}+d_{100}d_{001}+e_{100}e_{001})\rho_1\rho_3+
2(c_{010}c_{001}+d_{010}d_{001}+e_{010}e_{001})\rho_2\rho_3\biggr] \\
&& \\
\hat{\phi}_2&=&8m^2(d_{100}\rho_1+d_{010}\rho_2+d_{001}\rho_3)\biggl/\biggl[ 8m^2-\lambda[(c_{100}^2+d_{100}^2+e_{100}^2)\rho_1^2 \\
&&+(c_{010}^2+d_{010}^2+e_{010}^2)\rho_2^2+(c_{001}^2+d_{001}^2+e_{001}^2)\rho_3^2+2(c_{100}c_{010}+d_{100}d_{010}+e_{100}e_{010})\rho_1\rho_2 \\
&&+2(c_{100}c_{001}+d_{100}d_{001}+e_{100}e_{001})\rho_1\rho_3+
2(c_{010}c_{001}+d_{010}d_{001}+e_{010}e_{001})\rho_2\rho_3\biggr]\\
&& \\
\hat{\phi}_3&=&8m^2(e_{100}\rho_1+e_{010}\rho_2+e_{001}\rho_3)\biggl/\biggl[ 8m^2-\lambda[(c_{100}^2+d_{100}^2+e_{100}^2)\rho_1^2 \\
&&+(c_{010}^2+d_{010}^2+e_{010}^2)\rho_2^2+(c_{001}^2+d_{001}^2+e_{001}^2)\rho_3^2+2(c_{100}c_{010}+d_{100}d_{010}+e_{100}e_{010})\rho_1\rho_2 \\
&&+2(c_{100}c_{001}+d_{100}d_{001}+e_{100}e_{001})\rho_1\rho_3+
2(c_{010}c_{001}+d_{010}d_{001}+e_{010}e_{001})\rho_2\rho_3\biggr]
\end{eqnarray*}
\normalsize

Substituting the above expressions  into the system (\ref{system rho}) with $N_\phi=N_\rho=3$, we can check that these ansatz already are a exact solution for the linear sigma model.

\subsection{General case}\label{general case}

With the information we gathered in the previous sections, we can seek a solution for the general case with $N_\phi$ and $N_\rho$ arbitrary. Let us consider for a moment the following expressions:
\begin{equation}\label{general case 1}
\hat{\phi}_i={8m^2\sum_{j=1}^{N_\rho} c^{(i)}_{\delta_{1j},\delta_{2j},...,\delta_{N_\rho j}}\rho_j  \over 8m^2-\lambda\sum_{p=1}^{N_\phi} 
(\sum_{j=1}^{N_\rho}c^{(p)}_{\delta_{1j},\delta_{2j},...,\delta_{N_\rho j}}\rho_j)^2   },    \hspace{1 cm}     i=1,...,N_\phi
\end{equation}
\begin{equation}\label{general case 2}
 k_{i\mu}k_j^\mu+m^2=0,    \hspace{1 cm}    i,j=1,...,N_\rho
\end{equation}

On the one hand, if we substitute these expressions in the kinematic part of the model (\ref{system rho}), we have
\footnotesize
\begin{eqnarray*}
\sum_{p,q=1}^{N_{\rho}}k_{p\mu}k_q^\mu\biggl(\rho_p\rho_q\hat{\phi}_{i,\rho_p\rho_q}+\delta_{pq}\rho_p\hat{\phi}_{i,\rho_p}\biggr) &=& -m^2\biggl({8m^2\sum_{j=1}^{N_\rho}c_{\delta_{1j},...,\delta_{N_\rho j}}^{(i)}\rho_j\over 8m^2-\lambda\sum_{p=1}^{N_\phi} 
(\sum_{j=1}^{N_\rho}c^{(p)}_{\delta_{1j},\delta_{2j},...,\delta_{N_\rho j}}\rho_j)^2} \\
&+&  {8\lambda(8m^2)^2(\sum_{j=1}^{N_\rho}c_{\delta_{1j},...,\delta_{N_\rho j}}^{(i)}\rho_j)\sum_{p=1}^{N_\phi}(\sum_{j=1}^{N_\rho}c_{\delta_{1j},...,\delta_{N_\rho j}}^{(p)}\rho_j)^2\over [8m^2-\lambda\sum_{p=1}^{N_\phi} 
(\sum_{j=1}^{N_\rho}c^{(p)}_{\delta_{1j},\delta_{2j},...,\delta_{N_\rho j}}\rho_j)^2]^3}\biggr).
\end{eqnarray*}
\normalsize
On the other, if we substitute (\ref{general case 1}) on the potential part, we have
\small
\begin{eqnarray*}
m^2\hat{\phi}_i+\lambda\biggl(\hat{\phi}_i^3+\hat{\phi}_i\sum_{j=1}^{N_{\phi}-1}\hat{\phi}_{i+j}^2\biggr) &=& m^2\biggl({8m^2\sum_{j=1}^{N_\rho}c_{\delta_{1j},...,\delta_{N_\rho j}}^{(i)}\rho_j\over 8m^2-\lambda\sum_{p=1}^{N_\phi} 
(\sum_{j=1}^{N_\rho}c^{(p)}_{\delta_{1j},\delta_{2j},...,\delta_{N_\rho j}}\rho_j)^2} \\
&+&  {8\lambda(8m^2)^2(\sum_{j=1}^{N_\rho}c_{\delta_{1j},...,\delta_{N_\rho j}}^{(i)}\rho_j)\sum_{p=1}^{N_\phi}(\sum_{j=1}^{N_\rho}c_{\delta_{1j},...,\delta_{N_\rho j}}^{(p)}\rho_j)^2\over [8m^2-\lambda\sum_{p=1}^{N_\phi} 
(\sum_{j=1}^{N_\rho}c^{(p)}_{\delta_{1j},\delta_{2j},...,\delta_{N_\rho j}}\rho_j)^2]^3}\biggr).
\end{eqnarray*}
\normalsize

Therefore, we can easily check that (\ref{general case 1}) with condition (\ref{general case 2}) is a solution for the general case. Observe that we have ${1\over2}N_\rho(N_\rho+1)$ algebraic equations in (\ref{general case 2}) and $DN_\rho$ constantes $k_{i\mu}$ if we are in a D-dimensional space. Hence, for the system (\ref{general case 2}) be solvable, we need have $N_\rho\leq 2D-1$ different $\rho_j$.

\section{Graphical Illustration}\label{illustration}

In order to illustrate the analytical solution (\ref{general case 1}), let us consider the case $N_\phi=2$ and $N_\rho=4$ with
\begin{eqnarray*}
(k_{10},k_{11},k_{12},k_{13})&=&(2,-1,2,0)  \\
(k_{20},k_{21},k_{22},k_{23})&=&(-2,-1,-2,0)  \\
(k_{30},k_{31},k_{32},k_{33})&=&(1/2,-1,1/2,0)  \\
(k_{40},k_{41},k_{42},k_{43})&=&(-1/2,-1,-1/2,0).
\end{eqnarray*}
Substituting these parameters in (\ref{general case 1}) and rewriting the expression in the $x^\mu$ coordinates using the structure of exponential functions, we have
\begin{eqnarray}
\hat{\phi}_1&=&8m^2(c_{1000}^{(1)}e^{2t-x+2y}+c_{0100}^{(1)}e^{-2t-x-2y}+c_{0010}^{(1)}e^{{1\over 2}t-x+{1\over 2}y}+c_{0001}^{(1)}e^{-{1\over 2}t-x-{1\over 2}y})\biggl/  \nonumber  \\
&& \biggl[ 8m^2-\lambda[(c_{1000}^{(1)}e^{2t-x+2y}+c_{0100}^{(1)}e^{-2t-x-2y}+c_{0010}^{(1)}e^{{1\over 2}t-x+{1\over 2}y}+c_{0001}^{(1)}e^{-{1\over 2}t-x-{1\over 2}y})^2  \nonumber \\
&+&(c_{1000}^{(2)}e^{2t-x+2y}+c_{0100}^{(2)}e^{-2t-x-2y}+c_{0010}^{(2)}e^{{1\over 2}t-x+{1\over 2}y}+c_{0001}^{(2)}e^{-{1\over 2}t-x-{1\over 2}y})^2\biggr]               \label{phi1  rho 4}        \\
\hat{\phi}_2&=&8m^2(c_{1000}^{(2)}e^{2t-x+2y}+c_{0100}^{(2)}e^{-2t-x-2y}+c_{0010}^{(2)}e^{{1\over 2}t-x+{1\over 2}y}+c_{0001}^{(2)}e^{-{1\over 2}t-x-{1\over 2}y})\biggl/   \nonumber \\
&& \biggl[ 8m^2-\lambda[(c_{1000}^{(1)}e^{2t-x+2y}+c_{0100}^{(1)}e^{-2t-x-2y}+c_{0010}^{(1)}e^{{1\over 2}t-x+{1\over 2}y}+c_{0001}^{(1)}e^{-{1\over 2}t-x-{1\over 2}y})^2  \nonumber  \\
&+&(c_{1000}^{(2)}e^{2t-x+2y}+c_{0100}^{(2)}e^{-2t-x-2y}+c_{0010}^{(2)}e^{{1\over 2}t-x+{1\over 2}y}+c_{0001}^{(2)}e^{-{1\over 2}t-x-{1\over 2}y})^2\biggr]              \label{phi2  rho 4}
\end{eqnarray}

\begin{figure}
        \centering
        \begin{subfigure}[hbt]{0.25\textwidth}
                \includegraphics[width=\textwidth]{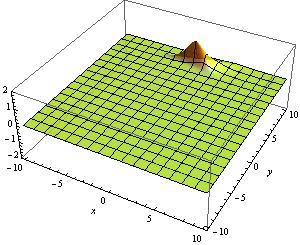}
                \caption{$\phi_1(t=-8,{\bf x})$}
                \label{fig vacuo}
        \end{subfigure}%
        \begin{subfigure}[hbt]{0.25\textwidth}
                \includegraphics[width=\textwidth]{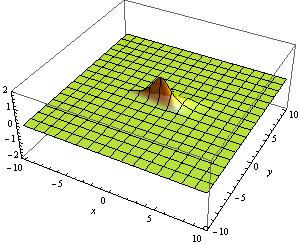}
                \caption{$\phi_1(t=0,{\bf x})$}
                \label{fig 1 sol}
        \end{subfigure}
        \begin{subfigure}[hbt]{0.25\textwidth}
                \includegraphics[width=\textwidth]{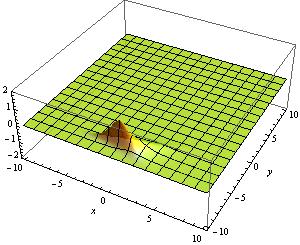}
                \caption{$\phi_1(t=8,{\bf x})$}
                \label{fig 2 sol}
        \end{subfigure}
        \begin{subfigure}[hbt]{0.25\textwidth}
                \includegraphics[width=\textwidth]{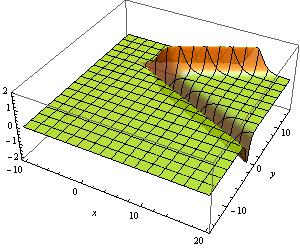}
                \caption{$\phi_2(t=-8,{\bf x})$}
                \label{fig vacuo}
        \end{subfigure}%
        \begin{subfigure}[hbt]{0.25\textwidth}
                \includegraphics[width=\textwidth]{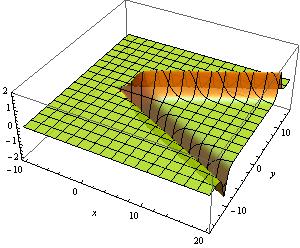}
                \caption{$\phi_2(t=0,{\bf x})$}
                \label{fig 1 sol}
        \end{subfigure}
        \begin{subfigure}[hbt]{0.25\textwidth}
                \includegraphics[width=\textwidth]{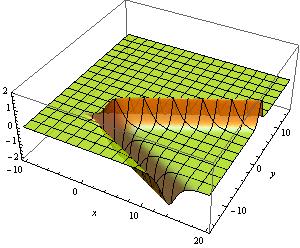}
                \caption{$\phi_2(t=8,{\bf x})$}
                \label{fig 2 sol}
        \end{subfigure}
        \caption{In these figures, we illustrate the solutions (\ref{phi1  rho 4}) and (\ref{phi2  rho 4}) for the linear sigma model with two fields, four traveling-wave and the parameters (\ref{par 1}) and (\ref{par 2})  in a succession of three graphics at different time for each field. The z-coordinate is omitted because we chose null components for $k_{j3}$.}
                \label{fig1}
\end{figure}

Observe that these solutions will present singularities if $\lambda$ is positive and the chosen values for $k_{i\mu}$ are real, therefore, we will consider only the case $\lambda<0$. For illustration purpose, let us consider $m=1$ and $\lambda=-1$. As we have chosen $k_{i3}=0$, we will display the solution as a sequence of three bi-dimensional graphics at different times. In figure \ref{fig1}, we show the graphic for the following combination of the arbitrary constants:
\begin{eqnarray}
(c_{1000}^{(1)},c_{0100}^{(1)},c_{0010}^{(1)},c_{0001}^{(1)})&=&\sqrt{8}(0,0,1,0)   \label{par 1}    \\
(c_{1000}^{(2)},c_{0100}^{(2)},c_{0010}^{(2)},c_{0001}^{(2)})&=&\sqrt{8}(1,-1,1,-1).   \label{par 2}
\end{eqnarray}

%$(c_{1000}^{(1)},c_{0100}^{(1)},c_{0010}^{(1)},c_{0001}^{(1)})=\sqrt{8}(0,0,1,0)$ and $(c_{1000}^{(2)},c_{0100}^{(2)},c_{0010}^{(2)},c_{0001}^{(2)})=\sqrt{8}(1,-1,1,-1)$ 

This is one of the many combinations we can obtain from the analytical solution, but it is interesting to notice a very localized pattern for $\phi_1$ produced by the combination of four traveling-waves that resembles a particle. A phenomenological study of this solution will be considered in a future work.

%%%%%%%%%%%%%%%%%%%%%%%%%%%%%%%%%%%%%%%%%%%%%%%%%%%%%%%%%%

\section{Efficiency in the processing of data}\label{comparison}

In this section, we will compare the algorithm used in this paper with the multiple exp-function method \cite{Ma1} in order to see the benefit of each method. The method in \cite{Ma1} was the first method to seek multi-wave solutions using a direct approach. On one hand, if we would like to use the multiple exp-method in the case of subsection \ref{subsection N2}, we would use the ansatz

\begin{equation}\label{multi exp}
\phi_i={\sum_{m,n=0}^M p_{i,m,n}\rho_1^m\rho_2^n \over\sum_{m,n=0}^N q_{i,m,n}\rho_1^m\rho_2^n},    \hspace{1.0 cm}   \partial_\mu\rho_{j}=k_{j\mu}\rho_j,    \hspace{1 cm}   i,j=1,2
\end{equation}
and the solution would be found for $M=1$ and $N=2$. Substituting this ansats in equation (\ref{system rho})  with $N_{\phi}=N_{\rho}=2$ and forming a system of algebraic equations with the coefficients of all powers of $\rho_k$, we have a system with 34 variables (4 $p_{i,m,n}$ and 9 $q_{i,m,n}$ for each field and 4 $k_{j\mu}$ for each wave). If we consider the case of subsection \ref{subsection N3} without any previous assumption, the number of variables jump to 117 (8 $p_{i,m,n}$ and 27 $q_{i,m,n}$ for each field and 4 $k_{j\mu}$ for each wave). Such systems can be quite complicated depending of the model.

On the other hand, the algorithm based on Padé approximants we have used splits the process of find the undetermined coefficients in two stages. Let us dig the calculation of the Taylor expansion in the case of subsection \ref{subsection N2}. The algebraic equation used to determine the firsts elements of the expansion are:
\begin{eqnarray*}
\hat{E}_{1;00} &=& c_{00}(m^2+\lambda(c_{00}^2+d_{00}^2))=0 \\
\hat{E}_{2;00} &=& d_{00}(m^2+\lambda(c_{00}^2+d_{00}^2))=0 \\
\hat{E}_{1;10} &=& 2\lambda c_{00}d_{00}d_{10}+c_{10}(k_{1\mu}k_1^\mu+m^2+\lambda(3c_{00}^2+d_{00}^2))       =0 \\
\hat{E}_{2;10} &=& 2\lambda c_{00}d_{00}c_{10}+d_{10}(k_{1\mu}k_1^\mu+m^2+\lambda(c_{00}^2+3d_{00}^2))       =0 \\
\hat{E}_{1;01} &=& 2\lambda c_{00}d_{00}d_{01}+c_{01}(k_{2\mu}k_2^\mu+m^2+\lambda(3c_{00}^2+d_{00}^2))       =0 \\
\hat{E}_{2;01} &=& 2\lambda c_{00}d_{00}c_{01}+d_{01}(k_{2\mu}k_2^\mu+m^2+\lambda(c_{00}^2+3d_{00}^2))       =0 
\end{eqnarray*}
In these system, we impose $c_{00}=d_{00}=0$ because we would like that the fields be null at infinite (similarly, this physical constraint could be imposed on ansats (\ref{multi exp}) for simplify the solution of the algebraic system).  So, the above system yields
\[
c_{00}=d_{00}=0,   \hspace{1 cm}   k_{1\mu}k_1^\mu+m^2=0,   \hspace{1 cm}   k_{2\mu}k_2^\mu+m^2=0,   \hspace{1 cm}   c_{10},c_{01},d_{10},d_{01}  \hspace{0.5 cm} \textnormal{arbitrary}  .
\]
For calculate the Padé approximants $[2/2]_\rho^{(i)}(\xi;{\cal S})$, we still need to find $c_{mn}$ and $d_{mn}$ for $m+n<=4$; however, the equations that yield these coefficients are linear and ease to solve. After we have calculated the Taylor expansion until the order we need, we stay with only 10 variables to be determined in the second stage of the algorithm ($c_{10}$, $c_{01}$, $d_{10}$, $d_{01}$ and 3 $k_{j\mu}$ for each wave).

Therefore, we can see that  the algorithm based on Padé approximants can organize and simplify the processing of data by transforming part of procedure in a linear system. However, this algorithm has a disadvantage. If we deal with a model that has a singularity at the origin in the $\rho_k$ variable, the multiple exp-method may perform better. The solution for such problem in the Padé approach is redefine the variables $\rho_k$ as
\[
\rho_j=e^{k_{j\mu}x^\mu} \hspace{0.2 cm} \to \hspace{0.2 cm} \rho_j=e^{k_{j\mu}x^\mu}-\alpha,  \hspace{0.5 cm}   \partial_\mu\rho_{j}=k_{j\mu}\rho_j  \hspace{0.2 cm} \to \hspace{0.2 cm}  \partial_\mu\rho_{j}=k_{j\mu}(\rho_j+\alpha),   \hspace{0.5 cm} \alpha=\textnormal{constant} 
\]
in order to avoid the singularity, but this could complicate the equation to be solved.

% using the procedure explained in section \ref{method}

%Besides the algorithm used in this paper, the multiple exp-function method \cite{Ma1}

\section{Conclusion}

In this paper, it was applied a method based on the Padé approximants in order to obtain traveling wave solutions for the linear sigma model. With the solutions found for two and three bosonic fields, we were able to write a solution for an arbitrary number of bosons and traveling waves.  The results of the current work show that the method developed in \cite{Ruy lambda phi 4} is robust and can be used to find explicit solutions in specific problems in classical field theory.

\section*{Acknowledgements}
I am thankful to D. Bazeia and L. Losano for discussions. The author thanks  CNPQ (402798/2015-5) for financial support.

\end{document}